\documentclass[english,aps,pra]{paper}
\usepackage[T1]{fontenc}
\usepackage[latin9]{inputenc}
\usepackage{bm}
\usepackage{amsmath}
\usepackage{amssymb}

\makeatletter
\newcommand{\lyxaddress}[1]{
\par {\raggedright #1
\vspace{1.4em}
\noindent\par}
}

\makeatother

\usepackage{babel}
\begin{document}

\title{Coherent functional expansions in quantum field theory}

\author{P. D. Drummond}
\maketitle

\lyxaddress{Centre for Quantum and Optical Science, Swinburne University of Technology,
Melbourne 3122, Australia\\
Kavli Institute for Theoretical Physics, UC Santa Barbara, USA. }
\begin{abstract}
A new formalism is introduced to treat problems in quantum field theory,
using coherent functional expansions rather than path integrals. The
basic results and identities of this approach are developed. In the
case of a Bose gas with point-contact interactions, this leads to
a soluble functional equation in the weak interaction limit, where
the perturbing term is part of the kinetic energy. This approach has
the potential to prevent the Dyson problem of divergence in perturbation
theory.
\end{abstract}

\section{Introduction}

One of the most difficult problems in theoretical physics is conceptually
very simple. How does one calculate the ground state of an interacting
many-body quantum system? Analytic solutions do not usually exist
in more than one space dimension, and a finite computer cannot easily
solve such exponentially hard problems \cite{Feynman_1982}. Conventional
path integrals \cite{Feynman1948RevModPhys.20.367}, are not exactly
soluble when there are interactions. Infinite order perturbation theory
with truncated sets of diagrams is frequently used to handle strongly
interacting systems. Approaches like this are being tested in experiments
in ultra-cold atomic physics \cite{RevModPhys.80.885,luo2007measurement}
using Feshbach resonances that can reach strongly interacting, universal
behavior \cite{ho2004universal,hu2007universal,hu2010universal} .

However, as first pointed out in the work of Dyson \cite{dyson1952divergence},
conventional perturbation theory does \emph{not} converge as a power
series in the coupling constant \cite{Suslov2005,Boyd1999}. This
cannot be cured using renormalization \cite{Jaffe1965}. Similar convergence
issues exist in other quantum theories \cite{Pernice1998PhysRevD.57.1144}.
They occur when the exact theory is not analytic at zero coupling;
an example is the quantum anharmonic oscillator \cite{zinn1996quantum}.
While other methods exist for Bose condensates \cite{GardinerPhysRevA.56.1414,CastinDumPhysRevA.57.3008,Morgan-gapless,SGardinerPhysRevA.75.043621},
any attempt to use the interaction term as a perturbation is likely
to have divergences for this reason \cite{Jaffe1965}. 

Here an approach is introduced in which coherent functional states
are used to treat a non-relativistic quantum field theory. This allows
one to treat the weakly interacting ground state in a Bose condensate,
where the interaction term is included in the \emph{unperturbed} part
of the problem.

The functional expansion used is an integral over amplitudes of coherent
states. Related stochastic techniques \cite{Carusotto:2001} and variational
methods \cite{He2012_QDynamics_FrontiersPhysics} are known. Here
we show that in the limit of weak interactions, one can define an
asymptotic ground state which includes the nonlinear coupling. This
leaves part of the kinetic energy as a perturbation, rather than the
interaction term. Identities and functional equations are obtained
in this Letter, together with limiting behaviour, and more detailed
properties will be treated elsewhere. 

\section{Coherent states and phase integrals\label{sec:Coherent-states-and}}

The notion of a coherent state has proved useful in understanding
the basis of coherence \cite{Schrodinger_CS,Glauber_1963_P-Rep}.
An unnormalized functional coherent state \cite{Bargmann:1961,Glauber_1963_P-Rep,graham1970functional,Steel1998}
of a bosonic quantum field $\hat{\Psi}\left(x\right)$ is
\begin{equation}
\left\Vert \alpha\right\rangle \equiv\exp\left(\int_{V}\alpha\cdot\hat{\Psi}^{\dagger}dx\right)\left|0\right\rangle ,\label{eq:Coherent-definition}
\end{equation}
with a coherent amplitude field $\alpha\left(x\right)$, where $\left|0\right\rangle $
is the vacuum state. The integration is over a space coordinate $x$
in a $D$ dimensional real vector space, volume $V$, with periodic
boundary conditions and volume measure $dx$. Arguments of functions
inside space integrals are omitted. Commutators for the scalar fields
are $\left[\hat{\Psi}\left(x\right),\hat{\Psi}^{\dagger}\left(x'\right)\right]=\delta^{D}\left(x-x'\right)\,$.
The coherent states are not orthogonal, since \cite{campbell1896law,baker,hausdorff1906symbolische}:
\begin{align}
\left\langle \alpha\right.\left\Vert \alpha'\right\rangle  & =e^{G\left[\alpha',\alpha^{*}\right]}\,,\label{eq:Bargmann-orthonormality}
\end{align}
where $G\left[\alpha,\alpha'\right]=\int_{V}\alpha^{*}\cdot\alpha'dx\,$.
Introducing a normalizing factor $G\equiv G\left[\alpha,\alpha\right]$,
the normalized functional coherent state is: $\left|\alpha\right\rangle \equiv\exp\left[-G/2\right]\left\Vert \alpha\right\rangle $.

Arbitrary quantum states can be written as functional expansions $\left|f\left[\alpha\right]\right\rangle $
and phase-space representations over coherent states \cite{Glauber_1963_P-Rep,Chaturvedi:1977,Drummond_1987_JOptSocAmB,Steel1998}.
These are defined here using the unnormalized coherent states, with
a functional measure $d[\alpha]$ such that:

\begin{equation}
\left|f\left[\alpha\right]\right\rangle _{\alpha}\equiv\int f\left[\alpha\right]\left\Vert \alpha\right\rangle d[\alpha]\,.
\end{equation}

With number conserving Hamiltonians, it is natural to use a basis
of number-projected coherent states, $\left\Vert \alpha\right\rangle _{N}$
\cite{DrummondReidArxiv2016}, which are obtained through projection
onto a particle number $N$ with a projection operator $\hat{P}_{N}$,
using a phase integral \cite{Imada_2007_GBMC}, 

\begin{equation}
\left\Vert \alpha\right\rangle _{N}\equiv\hat{P}_{N}\left\Vert \alpha\right\rangle =\int\frac{d\phi}{2\pi}e^{-i\phi N}\left\Vert \alpha e^{i\phi}\right\rangle .
\end{equation}
This is shown to be a number projection by a Taylor expansion of the
exponential in (\ref{eq:Coherent-definition}). As a result, 

\begin{equation}
\left\Vert \alpha\right\rangle _{N}\equiv\frac{1}{N!}\left(\int_{V}\alpha\cdot\hat{\Psi}^{\dagger}dx\right)^{N}\left|0\right\rangle .
\end{equation}

The inner product of $\left\Vert \alpha\right\rangle _{N}$ and $\left\Vert \alpha'\right\rangle _{N'}$
vanishes unless $N=N'$. One can similarly show that the inner product
of number-conserving coherent states is $\left\langle \alpha\right.\left\Vert \alpha'\right\rangle _{N}=G^{N}\left[\alpha,\alpha'\right]/N!$~,
by expanding the resulting exponential in a power series in $G\left[\alpha,\alpha'\right]$.

\subsection{Coherent phase fields}

It is useful to employ a phase field notation, such that $\alpha\left(x\right)=\exp\left(\theta\left(x\right)+i\phi\left(x\right)\right)$,
where $\phi\left(x\right)$, $\theta\left(x\right)$ are phase and
amplitude fields. Here $\theta\left(x\right)=\ln\left|\alpha\left(x\right)\right|=\ln r(x)$
determines the coherent logarithmic amplitude. A phase integral by
itself is known to give a complete expansion for any system with a
finite particle number per mode \cite{Drummond_Gardiner_PositivePRep},
so for simplicity we now assume that $r$ is a constant. The inner
products of two coherent functional states are
\begin{align}
\left\langle \alpha\right\Vert \left.\alpha'\right\rangle  & =\exp\left(\int_{V}rr'e^{i(\phi'-\phi)}dx\right)=K\left[\phi,\phi'\right]\,.
\end{align}
Similarly, the inner products for the number-projected case are: 
\begin{align}
\left\langle \alpha\right\Vert \left.\alpha'\right\rangle _{N} & =\frac{1}{N!}\left[\int_{V}rr'e^{i(\phi'-\phi)}dx\right]^{N}=K_{N}\left[\phi,\phi'\right]\,.
\end{align}

\subsection{Coherent phase functional states}

One can also write the quantum field state functional integral in
the form: 
\begin{equation}
\left|\psi\right\rangle \equiv\int\psi\left[\phi\right]\left\Vert \alpha\right\rangle d[\phi]\,,
\end{equation}
where the Jacobian for this measure change, $J=\left|d\left[\bm{\alpha}\right]/d[\phi]\right|$
, is absorbed into the definition of $\psi\left[\phi\right]$ . The
following inner product holds:
\begin{align}
\left\langle \psi\left|\right.\psi'\right\rangle  & =\int\int\psi^{*}[\phi]\psi'[\phi']K\left[\phi,\phi'\right]d[\phi]d[\phi']\,.
\end{align}

\section{\label{sec:Differential-identities}Differential identities}

Differential identities can be used to transform operator equations
into functional differential equations. Coherent states are eigenstates
of annihilation operators, so that $\hat{\psi}\left(x\right)\left\Vert \alpha\right\rangle =\alpha\left(x\right)\left\Vert \alpha\right\rangle \,.$
From differentiating Eq (\ref{eq:Coherent-definition}), one obtains
a differential identity for the creation operator:
\begin{eqnarray}
\hat{\psi}^{\dagger}\left(x\right)\left\Vert \alpha\right\rangle  & = & \frac{\delta}{\delta\alpha\left(x\right)}\left\Vert \alpha\right\rangle \,.
\end{eqnarray}
 For systems with a conserved particle number, the Hamiltonian generates
simple identities for number-conserving quadratic forms. Since the
global projection operator $\hat{P}_{N}$ commutes with any number-conserving
function of the field operators, the results below apply equally to
number-projected states. Applying these identities to the particle
density operator, $\hat{n}\left(x\right)=\hat{\psi}^{\dagger}\left(x\right)\hat{\psi}\left(x\right)$,
one obtains that:
\begin{eqnarray}
\hat{n}\left(x\right)\left\Vert \alpha\right\rangle  & = & \frac{1}{i}\frac{\delta}{\delta\phi\left(x\right)}\left\Vert \alpha\right\rangle \,.
\end{eqnarray}

For number-conserved states, $\hat{N}\left\Vert \alpha\right\rangle _{N}=N\left\Vert \alpha\right\rangle _{N}\,.$
In this case, it is useful to define a Fourier transform, $\phi\left(x\right)=\sum_{k}e^{ik\cdot x}\phi_{k}$,
so the functionals are now scalar functions of the vector $\bm{\phi}$.
On defining $\partial_{k}\equiv\partial/\partial\phi_{k}$, the corresponding
differential relationship is 
\begin{equation}
\frac{\delta}{\delta\phi\left(x\right)}=\frac{1}{V}\sum_{k}e^{-ik\cdot x}\partial_{k}\,.
\end{equation}
For the number density operator, one can introduce an \emph{average}
particle density $\rho=N/V$ in order to replace the derivative over
$\phi_{0}$, so that this can also be written as:
\begin{eqnarray}
\hat{n}\left(x\right)\left\Vert \alpha\right\rangle _{N} & = & \left[\rho+\frac{1}{iV}\sum_{k\neq0}e^{-ik\cdot x}\partial_{k}\right]\left\Vert \alpha\right\rangle _{N}\,.
\end{eqnarray}

\subsection{Functional identities}

To evaluate an observable or mean value property, consider an arbitrary
operator $\hat{O}$ so that:
\begin{equation}
\hat{O}\left|\psi\right\rangle \equiv\int\psi\left[\phi\right]\hat{O}\left\Vert \alpha\right\rangle d[\phi]\,.
\end{equation}
Any operator $\hat{O}$ can be transformed so that $\hat{O}\left|\psi\right\rangle \equiv\int\left\Vert \alpha\right\rangle \mathcal{O}\left[\phi\right]\psi\left[\phi\right]d[\phi]\,,$where
$\mathcal{O}$ is a differential operator acting on $\psi$. Hence,
provided $\left|\psi\right\rangle $ is normalized, the expectation
of an operator is: 
\begin{align}
Tr\left(\hat{\rho}\hat{O}\right) & =\int\int\psi^{*}[\phi]K\left[\phi,\phi'\right]\mathcal{O}\left[\phi'\right]\psi'[\phi']d[\phi]d[\phi']\,.
\end{align}

The identities are obtained as follows. Firstly, the action of annihilation
operators is simply:
\begin{equation}
\hat{\psi}\left(x\right)\left|\psi\right\rangle =\int\psi\left[\phi\right]\alpha\left(x\right)\left\Vert \alpha\right\rangle d[\phi]\,,
\end{equation}
and for creation operators, after partial integration,
\begin{eqnarray}
\hat{\psi}^{\dagger}\left(x\right)\left|\psi\right\rangle  & = & \int\left\Vert \alpha\right\rangle \frac{i}{\alpha\left(x\right)}\frac{\delta\psi\left[\phi\right]}{\delta\phi\left(x\right)}d[\phi]\,.
\end{eqnarray}

This can be iterated for any polynomial in field operators. For example,
the particle density operator gives, after partial integration, 
\begin{eqnarray}
\hat{n}\left(x\right)\left|\psi\right\rangle  & = & i\int\frac{\delta\psi}{\delta\phi\left(x\right)}\left\Vert \alpha\right\rangle d[\phi]\,.
\end{eqnarray}
These differential identities are also applicable to the number projected
coherent phase states, apart from the global phase derivative. This
can be removed using the mapping $\partial_{0}\rightarrow\{-iN\}$,
after partial integration. Terms like this will be added in braces
to indicate a replacement can be made. Thus, the identities apply
to either $\psi$ or to $\psi_{N}$, together with the substitution
$\mathcal{O}\rightarrow\mathcal{O}_{N}$, $K\rightarrow K_{N}$ . 

\section{Hamiltonian as a differential operator\label{sec:Hamiltonian-as-a}}

Next, consider a non-relativistic interacting Bose gas with a Hamiltonian
of form $\hat{H}=\hat{H}^{K}+\hat{H}^{U}+\hat{H}^{I}\,,$ where the
energy includes a chemical potential \cite{fetter2003quantum}, so
that
\begin{align}
\hat{H}^{K} & =-\frac{\hbar^{2}}{2m}\int\hat{\psi}^{\dagger}\nabla^{2}\hat{\psi}dx\,,\nonumber \\
\hat{H}^{U} & =\int\left(U-\mu\right)\hat{n}dx\,,\nonumber \\
\hat{H}^{I} & =\int\int\hat{n}\left(x\right)U^{I}\left(x-x'\right)\hat{n}\left(x'\right)dxdx'\,.
\end{align}

The ground state or energy eigenstate is a number eigenstate $\left|\psi\right\rangle _{N}$,
since $\hat{N}$ and $\hat{H}$ commute. It is assumed that $U^{I}>0$,
in order to have a stable, bounded Hamiltonian. The Hilbert space
operators are now expressed using differential identities so that:
\begin{equation}
\hat{H}\left|\psi\right\rangle =-\frac{\hbar^{2}}{2m}\int\left\Vert \alpha\right\rangle \mathcal{L}\left[\phi\right]\psi\left[\phi\right]d[\phi]\,,
\end{equation}
 where differential operators act on all terms to the right including
the wave-function $\psi$, and the ground state of $\hat{H}$ has
the largest eigenvalue of $\mathcal{L}$.

\subsection{Differential identities for the Hamiltonian}

For the kinetic energy term, the coherent eigenvalue identity for
the annihilation operator gives 
\begin{align}
\hat{H}^{K}\left|\psi\right\rangle  & =\frac{\hbar^{2}}{2m}\int\left\Vert \alpha\right\rangle \left\{ \int\left[\frac{1}{i\alpha}\nabla^{2}\alpha\right]\frac{\delta}{\delta\phi}dx\right\} \psi\left[\phi\right]d[\phi]\,.
\end{align}
 If the coherent amplitude is chosen as constant in space then the
following choice is possible, although it is not the only one:
\begin{eqnarray}
\mathcal{L}^{K} & = & \int\frac{\delta}{\delta\phi}\left[\nabla^{2}\phi+i\left|\nabla\phi\right|^{2}\right]dx\,.
\end{eqnarray}

To obtain an identity for an external potential energy term, set $u(x)=2m\left(U(x)-\mu\right)/\hbar^{2}$,
and the simplest differential operator is:
\begin{align}
\mathcal{L}^{U}\left[\phi\right] & =-i\int u\frac{\delta}{\delta\phi}\psi\left[\phi\right]dx\,,
\end{align}
where as elsewhere, the position arguments of $u(x)$ are suppressed
inside a space integral. In the case of an external potential with
Fourier transform
\begin{equation}
u\left(x\right)=\sum_{k}u_{k}e^{ikx}\,,
\end{equation}
the differential operator for the corresponding energy term in the
general case is $\mathcal{L}^{U}\left[\phi\right]=-i\sum_{k}u_{k}\partial_{k}\psi\,,$
and for a number conserving expansion:
\begin{equation}
\mathcal{L}_{N}^{U}\left[\phi\right]=-\left[Nu_{0}+i\sum_{k\neq0}u_{k}\partial_{k}\psi\right]\,.
\end{equation}

Defining $U^{I}(x)=\sum U_{k}^{I}\exp(ikx)$, the interaction Hamiltonian
becomes:
\begin{eqnarray}
\hat{H}^{I}\left|\psi\right\rangle  & = & -\int\left\Vert \alpha\right\rangle \int\int U^{I}(x-x')\frac{\delta^{2}\psi\left[\phi\right]}{\delta\phi\left(x\right)\delta\phi\left(x'\right)}dxdx'd[\phi]\,\nonumber \\
 & = & -\frac{g}{V}\left\Vert \alpha\right\rangle \left[\sum_{k}U_{k}^{I}\partial_{k}\partial_{-k}-\left\{ N^{2}U_{0}^{I}\right\} \right]\psi d[\phi]\,.
\end{eqnarray}
Setting $\gamma_{k}=2mU_{k}^{I}/\left(\hbar^{2}V\right)$, the differential
operator for the interaction term in the general case is:
\begin{align}
\mathcal{L}^{I} & =V\int\int\gamma(x-x')\frac{\delta^{2}\psi\left[\phi\right]}{\delta\phi\left(x\right)\delta\phi\left(x'\right)}dxdx'\,=\left[\sum_{k}\gamma_{k}\partial_{k}\partial_{-k}-\left\{ \gamma_{0}N^{2}\right\} \right]\psi\,.
\end{align}

\subsection{Overall result}

The quantum field evolution can now be mapped into a differential
equation with the form of a functional Fokker-Planck equation, such
that:
\begin{equation}
\frac{2m}{i\hbar}\frac{\partial\psi}{\partial t}=-\mathcal{L}\psi\,.
\end{equation}
This has the standard form of multi-variate Fokker-Planck linear operators,
apart from the imaginary term. One can write this in Fourier components
as
\begin{align}
\mathcal{L} & =\sum_{k}\left[-\partial_{k}A_{k}+\gamma\partial_{k}\partial_{-k}\right]-(\bar{e}_{N})\,,
\end{align}
where in the number conserved case the $k=0$ term is omitted, and
the term $\bar{e}_{N}=N\left(u_{0}+\gamma N\right)$ is included instead.
The first order drift term is:
\begin{align}
A_{k} & =-k^{2}\phi_{k}+i\left(u_{k}-\sum_{q}q\cdot\left(k-q\right)\phi_{q}\phi_{k-q}\right)\,.
\end{align}
For simplicity, we now assume an effective delta-function potential
with a momentum cutoff, so $\gamma_{k}=\gamma$, giving a functional
Fokker-Plank operator in phase:
\begin{align}
\mathcal{L} & =\int\left[\frac{\delta}{\delta\phi}\left(\nabla^{2}\phi+i\left[\left|\nabla\phi\right|^{2}-u\right]\right)+\gamma V\frac{\delta^{2}}{\delta\phi^{2}}\right]dx\,.
\end{align}

\section{Fokker-Planck equation}

In this section, the functional identities for the Hamiltonian are
assembled and put to work. The goal is to understand the equivalent
of the free-particle eigenstates. The limit of weak, but nonzero interactions
is especially straightforward. 

\subsection{Functional eigenvalue problem}

Using the identities derived above to write the energy eigenvalue
problem in a functional form gives:
\begin{eqnarray}
\hat{H}\left|\psi\right\rangle  & = & E\left|\psi\right\rangle =-\frac{\hbar^{2}}{2m}\int\mathcal{L}\left[\phi\right]\psi\left[\phi\right]\left\Vert \alpha\right\rangle d[\phi]\,.
\end{eqnarray}

Defining the eigenvalue as $E=\hbar^{2}e/\left(2m\right)$  the quantum
field eigenvalue problem can be mapped into a functional calculus
problem. It is simplest to define a chemical potential to shift the
energy origin, so that the ground state has eigenvalue $e=0$, which
defines an implicit condition on $\mu$. In the homogeneous case this
has the general form of a Fokker-Planck equation, $\mathcal{L}\psi=\left[-\partial_{k}A_{k}+\gamma\partial_{k}\partial_{-k}\right]\psi=0\,.$

 From the physics viewpoint, the meaning of the two terms is clear.
The diffusion term, proportional to $\gamma$, is the coupling term
which originates in the nonlinear interaction between the bosons.
It results in noise-like diffusive behaviour. The drift terms $A_{k}$
originate in the single-particle kinetic and potential energy. 

\subsection{Scaling behavior}

It is important to understand how the solutions behave in different
limits, by introducing a scaled coupling $\gamma'=\kappa\gamma$,
and a corresponding Hamiltonian $\hat{H}'$, where $\gamma$ is now
used to define a reference interaction strength. The occupation number
$N'$ is varied such that $N'=N/\kappa^{p}$, and the potential $u$
is taken to be uniform, such that $u'=\kappa^{q}u$. Here $p,q$ are
chosen so that so that the limit is physically nontrivial, while $N$
is a reference particle number. 

For $\psi\left(\phi\right)$ to have a well-defined limit, define
a scaled phase variable $\phi\equiv\phi^{(0)}/\kappa^{p}$, relative
to an unscaled phase $\phi^{(0)}$. The total number of field degrees
of freedom is held constant, and the changed functional measure is
absorbed into the definition of $\psi$. The notation $\left|\psi\right\rangle '$
now denotes a quantum state with a phase rescaling, where $\left|\psi\right\rangle '\equiv\int\psi\left[\phi\right]\left|\phi\kappa^{p}\right\rangle d[\phi]\,.$
From the calculus chain rule,
\begin{equation}
\frac{\partial}{\partial\phi_{k}^{(0)}}=\frac{\partial}{\kappa^{p}\partial\phi{}_{k}}\,.
\end{equation}

Therefore, in the new coordinates and fields, $\hat{H}'\left|\psi\right\rangle '=\int\mathcal{L}'\left[\phi\right]\psi'\left[\phi\right]\left|\phi\kappa^{p}\right\rangle d[\phi]\,,$
where the rescaled differential operator is now

\begin{eqnarray}
\mathcal{L}' & = & \int\frac{\delta}{\delta\phi}\left[\nabla{}^{2}\phi+i\left(\kappa^{p}\left|\nabla\phi\right|^{2}+u\kappa^{q-p}\right)+\gamma V\kappa^{1-2p}\frac{\delta}{\delta\phi}\right]dx\,.
\end{eqnarray}

\subsection{Weak coupling limit}

Consider a phase scaling transformation such that $\phi\equiv\phi^{(0)}/\sqrt{\kappa}$,
and $p=q=1/2$, which gives: 
\begin{equation}
\mathcal{L}'=\int\frac{\delta}{\delta\phi}\left[\nabla{}^{2}\phi+i\left(\epsilon\left|\nabla\phi\right|^{2}+u\right)+\gamma V\frac{\delta}{\delta\phi}\right]dx\,.
\end{equation}
Here $\epsilon=\sqrt{\kappa}$, so for small $\kappa\rightarrow0$,
the correlated term in $\left|\nabla\phi\right|^{2}$ is negligible.
Neglecting this, the Hamiltonian has a universal form, but it still
includes the coupling term. Setting $\mathcal{L}'\approx\mathcal{L}^{W}$
in this limit, the differential operator is:
\begin{equation}
\mathcal{L}^{W}=\int\frac{\delta}{\delta\phi}\left[\nabla{}^{2}\phi+iu+\gamma V\frac{\delta}{\delta\phi}\right]dx\,.
\end{equation}

With number conserving expansions, this can be rewritten in terms
of momentum expansions as:

\begin{align}
\mathcal{L}_{N}^{W} & =\sum_{k\neq0}\frac{\partial}{\partial\phi_{k}}\left\{ k^{2}\phi{}_{k}+\gamma\frac{\partial}{\partial\phi{}_{k}^{*}}\right\} -N\left(u_{0}+\gamma N\right)\,.\label{eq:Number-conserving weakly interacting H}
\end{align}
The ground state, number conserving solution of $\mathcal{L}_{N}^{W}\psi_{g}=0$
is a Gaussian in phase, with increasing phase-fluctuations at long
wavelength, of form: 
\begin{equation}
\psi_{g}\left(\phi\right)=\exp\left[-\frac{1}{2\gamma}\left(\sum_{k}k^{2}\left|\phi{}_{k}\right|^{2}\right)\right]\,.
\end{equation}
The corresponding physical energy per particle is $E/N=g\rho$, where
$\rho=N/V$ is the mean particle density. This gives the usual mean-field
energy in the absence of the kinetic correlation term, as one would
physically expect for this limit of weak couplings and high density. 

A crucial issue is whether the unscaled term $i\epsilon\left|\nabla\phi\right|^{2}$,
with $\epsilon=1$, can be used in a perturbation expansion around
$\epsilon=0$. In QED, Dyson showed that perturbation theory cannot
converge because the the exact theory is not analytic. It is discontinuous
under a sign change in the charge. Here a sign change in $\epsilon$
cannot cause a discontinuity. It corresponds to conjugation with $\mathcal{L}\rightarrow\mathcal{L}^{*}$,
which has no effect on the eigenvalue for a constant potential, and
simply conjugates the eigenfunction. This is not yet a proof of analyticity
or convergence, but it does show that instability under sign change
of the perturbation is not present with this approach. While the differential
operator $\mathcal{L}$ is not hermitian, this does not prevent the
use of perturbation theory \cite{Sternheim1972}. 

\section{Summary}

Here the weakly interacting ground state is the exact solution to
a part of the Hamiltonian in differential notation, which already
includes interactions. In this approach, the additional term required
to obtain the full Hamiltonian is now a part of the kinetic energy.
The perturbation does not include interactions, and there are no instabilities
under a sign change of the perturbation term. As a result, this type
of approach has the potential for a perturbation theory without Dyson's
dilemma of nonconvergence.

\section*{Acknowledgements}

This research was supported in part by the National Science Foundation
under Grant No. NSF PHY-1125915, and by the Australian Research Council.

\bibliographystyle{iopart-num}
\bibliography{QFT_references}

\end{document}